\documentclass[12pt,reprint]{aastex}
\usepackage{natbib}
\slugcomment{\large \bf Accepted for publication in The~Astrophysical~Journal~Letters}
\shorttitle{Dust Scattering Polarization of Exoplanet}
\shortauthors{Sengupta}
\begin{document}
\title{Cloudy Atmosphere of the Extra-solar Planet HD189733b : A Possible Explanation of the 
Detected B-band Polarization}
\author{Sujan Sengupta\altaffilmark{1,2}}
\altaffiltext{1}{Indian Institute of Astrophysics, Koramangala, Bangalore 560034, India}
\altaffiltext{2}{TIARA-ASIAA/National Tsing Hua University, Hsinchu, Taiwan;\\
E-mail: sujan@iiap.res.in}
\begin{abstract}
The peak amplitude of linear polarization detected recently from an extrasolar hot
giant planet HD 189733b, is a few times of $10^{-4}$, more than an order of magnitude
higher than all theoretical predictions. Rayleigh scattering off $H_2$ and $He$ may 
although give rise to a planet-star flux ratio of the order of $10^{-4}$ in the blue band,
it cannot account for the high polarization unless the planet has an unusually extended
atmosphere. Therefore, it is suggested that the high polarization should be attributed
to the presence of a thin cloud of sub-micron size dust grains in the upper visible
atmosphere which supports the observational finding of an almost feature-less transmission 
spectrum in the optical with no indication of the expected alkaline absorption features.
It is found that the polarimetry observation allows for a small eccentricity of the orbit
that is predicted from the time delay of the secondary eclipse of the planet. The estimated
longitude of the ascending node is $16^o\pm6$ which interestingly coincides with the 
observationally inferred location of the peak hemisphere-integrated brightness. 
\end{abstract}

\section{Introduction}

Polarization has always been an efficient tool to probe the physical
properties in the environment of various astrophysical objects.
\cite{sen01} predicted detectable amount of
linear polarization from L dwarfs because of the presence of condensates in the
visible atmosphere. Subsequently, linear polarization attributed
to dust scattering \citep{sen03,sen05} has been detected in several brown dwarfs
\citep{menard,osorio} whose atmospheres resemble those of irradiated extrasolar planets,
the 'Hot Jupiters'.  The use of polarimetry in detecting and understanding
the physical properties of extra-solar planets, especially the close-in-planets or the
so called roaster such as the first discovered extra-solar planet 51 Peg b \citep{mayor95},
is emphasized by \cite{seager00,seager03,sen06}. 

 Recently, \cite{berd} have reported detection of linear polarization in the blue band from
one of the well-studied 'Hot Jupiter' HD 189733b \citep{bouchy}. If the observation is 
correct, then it should have several important implications to our understanding of the 
physical processes of the planetary atmosphere. The observed peak amplitude of polarization
is about $10^{-4}$ implying the B-band flux ratio of the planet and the star to be more
than one order of magnitude higher than that predicted by current theroetical models
\citep{showman,burr07,barman}. 
 In order to model and interpreate the observed polarization, \cite{berd} have assumed
 the planet as a Lambert sphere of perfectly reflecting surface.
 A Lambert sphere follows Lambert's law of diffuse reflection. On this law the
diffusely reflected light is isotropic in the outward hemisphere and is natural,
independently of the state of polariation and the angle of incident light 
(Chandrasekhar 1960, p. 147).  That a Lambert sphere of perfectly reflecting surface yields
 zero polarization is also shown by \cite{stam06}. 
Consequently, the model fit and the inference by \cite{berd} that the
observed polarization implies that the planet has an abnormally large scattering
radius of 1.5-1.7 $R_J$ is unphysical.

  In the present letter, I show that single scattering of photons by sub-micron size grains in
a thin layer of silicate cloud in the upper atmosphere of HD 189733b may explain the observed
polarization. I discuss the polarization model in the next section. The cloud model adopted
is discussed in section~3. The results are 
discussed in the fourth section and specific conclusions are made in the last section.

\section{Theroetical Model }\label{obl}

The state of polarization of light is described by the Stoke parameters, i, q, u and v.
The parameter i is the total scalar specific intensity of radiation.
It is the complete flux of radiant energy inside the unit intervals of
frequency, time, solid angle, and area perpendicular to the flux.
This flux includes all radiation independently on polarization.
Polarization is described by the parameters q, u, v. These parameters
are proportional to the scalar specific intensity and have the same dimension.
For linear polarization, v=0.
On the other hand, in a plane-parallel scattering medium,
u is zero \citep{chandra60}. The amount of polarization is
 defined by the ratio of the polarized intensities and the total intensity. For an unresolved
 extrasolar planet, the total intensity is the sum of the total reflected intensity and the 
unpolarized stellar intensity.
Here,  I define the polarization as the normalized Stokes parameters Q and U  in a scattering 
reference plane through the centres of the star, the planet and the observer. The reference
 plane can be transformed to another one by using the Mueller rotation matrix
 \citep{chandra60} .
 Since the reflected intensity is negligible as compared to the stellar intensity, my
 normalization is with respect to the stellar intensity only. The Stokes parameters are 
integrated over the planetary disk. 
I assume single scattering which simplify the model calculations.
Since the dust density is presumed to be low and scattering by atoms and molecules does
not contribute to polarization significantly, single scattering approximation
is reasonable for the region where the optical depth $\tau < 1$. In the present model,
we incorporate a sufficiently thin cloud layer located between 0.2 and 0.1 bar of pressure
level (see section~\ref{aa}).
If present, multiple scattering can reduce the degree of polarization by a
few orders of magnitude \citep{sen01} because the planes
of the scattering events are randomly oriented and average each other's
contribution out from the final polarization.
The model, described in details in \cite{sen06}, is based on the formalism given in \cite{sim}.
 In a circular orbit, the normalized Stokes
parameter $Q$ and $U$ are given as a harmonic series :
\begin{equation}
Q(k,i,\Lambda)=\sum^{\infty}_{m=0}[p_m(k,i)\cos m\Lambda+q_m(k,i)\sin m\Lambda]
\end{equation}
\begin{equation}
U(k,i,\Lambda)=\sum^{\infty}_{m=0}[u_m(k,i)\cos m\Lambda+v_m(k,i)\sin m\Lambda]
\end{equation}
where $k=2\pi/\lambda$, $\lambda$ being the wavelength, $i$ is the orbital
inclination angle and $\Lambda$ is the orbital phase angle. The harmonic co-efficients
 are given by
\begin{equation}
\left(\begin{array}{c} p_m \\ q_m \end{array}\right)=\frac{2\pi}{k^2}
\sum^{\infty}_{l=M}F_{l2}(k)G^l_m(i)\left(\begin{array}{c} \eta_{lm} \\
\xi_{lm}\end{array}\right), m=0,1,2,3,\cdots
\end{equation}
\begin{equation}
\left(\begin{array}{c} u_m \\ v_m \end{array}\right)=\frac{2\pi}{k^2}
\sum^{\infty}_{l=M}F_{l2}(k)H^l_m(i)\left(\begin{array}{c} -\xi_{lm} \\
\eta_{lm}\end{array}\right), m=0,1,2,3,\cdots
\end{equation}
$M=max(2,m)$ and $G^l_m(i)$, $H^l_m(i)$ are given in \cite{sim}.
$\eta_{lm}$ and $\xi_{lm}$ are related with the density distribution in the
co-rotating frame and are given by
\begin{equation}\label{qq}
\left(\begin{array}{c} \eta_{lm} \\ \xi_{lm}\end{array}\right)=
\left[\frac{(2l+1)(l-m)\!}{4\pi(l+m)\!}\right]^{1/2}\int n'(r,\theta,\phi)P^m_l(\cos\theta_i)\left(\begin{array}{c}
\cos m\phi \\ sin m\phi\end{array}\right)\sin\theta d\theta d\phi dr.
\end{equation}
 where $n'(r,\theta,\phi)$ is the number density of scatterer in the co-rotating
frame, $\theta_i$ is the viewing angle and
$P^m_l$ is the associated Legendre function of the first kind.
$F_{l2}(k)$ is related to the scattering function and is given in \cite{sim,sen06}. The effect
of the optical properties, the shape and size of dust grains are incoporated through this
function. In the present work we consider spherical dust particles as scatterer.

  I assume an ellipsoidal distribution 
of scatterers illuminated by an unpolarized and point-like light source such that 
\begin{eqnarray}\label{abc3}
\eta_{lm}=2\pi\left[\frac{(2l+1)(l-m)\!}{4\pi(l+m)\!}\right]^{1/2}P^m_l(0) \int^{R_1}_{R_2}n(r)dr
\int^{1}_{-1}\frac{P_l(\mu)d\mu}{[1+(A^2-1)\mu^2]^{1/2}},
\end{eqnarray}
where $R_1$ and $R_2$ are the outer and the inner equatorial axis length of the
oblate planetary atmosphere, $A$ is the ratio of the length of the equatorial axis
to the polar axis such that the oblateness $f=1-1/A$.
and $\mu=\cos\theta$. Multi-poles up to l=5 and up to fifth harmonic,
i.e., m=0,1,2,3,4,5 are taken.

  If the light source is assumed to be 
point-like, the incident specific intensity from the star becomes distance independent.
 Hence, the amount of polarization or the normalized Stokes parameters  $Q$
 and $U$ becomes distance independent and I calculate them directly assuming spherical 
scatterers. It's worth mentioning here that the model does not calculate the reflected flux
from the planet but it estimates the amount of polarization from the unresolved system as seen
edge-on at an inclinatioin $i=90^o$ so that $\xi_{lm}=0$ in equation~\ref{qq}.
 However, the distribution and physical properties of scatterers depend on the thermal
 structure of the planetary atmosphere which is dependent on the orbital separation and on the
 amount of the stellar flux. The observable Q and U are obtained by rotating the scattering
 reference plane by an angle $\Omega$, the longitude of the ascending node of the planet. 

 For a slow rotator, the relationship for the oblateness $f$ of a stable
polytropic gas configuration under hydrostatic equilibrium is given by \cite{cha33}
as $f=2C\Omega^2R^3_e/(3GM)$,
where $M$ is the total mass, $R_e$ is the equatorial radius and $\Omega$ is
the angular velocity
of the object. $C$ is a constant whose value depends on the polytropic index.
For a polytropic index of $n=1.0$, $C=1.1399$, which is appropriate for
Jupiter \citep{hub84}.  \cite{bar03} modeled
the planet HD209458b and estimated its oblateness to be about 0.00285 whereas
the polytropic approximation yields a value of 0.00296.
Considering HD 189733b is tidally locked with its parent star, we estimate its spin-induced
oblateness to be about 0.003 by taking the orbital period of 2.218 days \citep{bouchy,bakos,
winn07}, radius $1.15R_J$, surface gravity 1995.0 $cm s^{-2}$ and assuming
 a polytropic equation of state with the polytropic index n=1 for the density distribution
 of the entire planet.

\section{Dust Distribution and Location}\label{aa}

The dust distribution in the atmosphere is calculated based on the one
dimensional cloud model of \cite{coo03}. The number density of cloud particles is given by
$n(P)=3q_c\rho\mu_d/(4\pi r^3\mu\rho_d)$
where $\rho$ is the mass density of the surrounding gas, $r$ is the cloud
particle radius, $\rho_{d}$ is the mass density of the dust condensates,
$\mu$ and $\mu_d$ are the mean molecular
weight of atmospheric gas and condensates respectively.
The condensate mixing number ratio ($q_c$) is given as
$q_c=q_{below}P_{c,l}/P$
for heterogeneously condensing clouds where $q_{below}$ is the fraction of
condensible vapor just below the cloud base, $P_{c,l}$ is the pressure at
the condensation point, and $P$ is the gas pressure in the atmosphere.
Given the equilibrium temperature of HD 189733b, condensates, if formed,
should be dominated by silicates in the form of forsterite or silicate oxide as it is
the case for L dwarfs of similar effective temperature \citep{hel08}. 
The values of $\mu_d$, $\rho_{d}$ and $q_{below}$
for forsterite  are taken from \cite{coo03}. I have adopted the dust-free
temperature-pressure profile of HD 189733b with the day-night heat distribution
parameter $P_n=0.3$ \citep{burr07}. This is kindly provided  by Adam Burrows 
(private communication).  
The location of the cloud base for different atmospheric models
and different chemical species is determined by
the intersection of the T-P profile of the atmosphere model and the
condensation curve $P_{c,l}$ as prescribed in \cite{coo03}.
Taking the condensation curve for forsterite as given in
\cite{sudars03}, we determine the base of the cloud from the T-P profiles
of  HD 189733b to be at 0.2 bar of pressure height. I consider the deck of the cloud
at 0.1 bar pressure level. This makes the cloud sufficiently thin so that
single scattering by dust grains is favored.

\section{Results and Discussions}
 
As mentioned in section~1, \cite{berd} obtained a fit of the observed polarimetric data
by assumeing a perfectly reflecting Lambert sphere with the radius as large as 1.5-1.7
$R_J$. While a Lambert sphere does not yield any polarization, the argument in favour of 
an unusually large Rayleigh scattering exosphere is somewhat prematured at this stage.
 Taking the scattering radius to be the same as the optical radius measured through
transit photometry, the geometric albedo as implied by the polarization is larger
than 2/3 which is the geometric albedo of a Lambert sphere of perfectly reflecting
surface. This may not be impossible as some Solar system objects have geometric albedo
exceeding unity due to strong back-scattering. However, \cite{berd} interpreated the
polarization by considering smaller albedo but an abnormally large planetary radius.
Recent observation of the exoplanet TrES-3 by \cite{winn08} does not
support such interpretation. These authors find upper limits on the planet's geometric
albedo in the $i,z$, and $R$ bands as
0.30, 0.62 and 1.07 respectively. Thus they rule out the presence of highly reflective
clouds in the atmosphere of TrES-3. It is worth mentioning that the geometric albedo and
the degree of polarization are two distinct physical quantities. The geometric albedo
is calculated from the ratio of the incident starlight to the sum of the isotropic
and anisotropic components of the emergent radiation from the planet. On the other hand, 
the degree of polarization from an unresolved planet is the ratio of the emergent
anisotropic radiation from the planet to the star light. While albedo is always non-zero,
 polarization could be zero if there is no anisotropy in the emergent planetary radiation.
 Therefore, the photometric
study by \cite{winn08} implies that the degree of polarization of TrES-3 should be very small
although it can have high albedo.  However, the conclusion for TrES-3 may not be applicable to
other planets, and in particular to HD 189733b. In fact, \cite{pont} have reported an almost
featureless transmission spectrum between 550 and 1050 nm with no indication of the expected alkaline 
absorption features which suggests the presence of a haze of sub-micron particles in the
upper atmosphere of HD 189733b.  \cite{burr08}  have found that a
high star-to-planet flux ratio in the blue is possible due to Rayleigh scattering off
$H_2$ and He. This although explains a high geometric albedo implied by the observed
polarization, it cannot explain the peak amplitude of the polarization itself unless the
scattering radius is much larger than that infered from transit observation.

   Absorption in the stellar Lyman $\alpha$ line observed during the transit of the
extrasolar planet HD 209458b by \cite{vidal03} 
revealed high velocity atomic hydrogen at great distances from the planet. This
is interpreated by \cite{vidal04} as hydrogen
atoms escaping from an extended atmosphere of the planet that is possibly undergoing
hydrodynamic blow-off. This interpretation however, fails to get theoretical support
\citep{hub07} and lead to controversy \citep{ben07}.
Recently, \cite{holm08} have provided
a viable alternative interpretation of the measured transit-associated Lyman alpha
absorption as the interaction between the exosphere of HD 209458b and the stellar
wind. These authors suggest a slow and hot stellar wind near the planet at the time
of observation. This interpretation is consistent with the energetic neutral atoms
around Solar System planets that is observed to form from charge exchange between solar
wind protons and neutral hydrogen from the planetary exospheres. Under such situation and
in the absence of any signature obtained by any other method, it would highly be premature
to consider an extended exosphere of HD 189733b in order to explain the observed
polarimetric data.
 
The re-binned polarization data shows negative value of Q at all phase angles. Because of
the large error-bars in U, I choose the value of U corresponding to the maximum value of Q
and find that $U/Q=0.63\pm0.3$. Hence, for a small value of U in the scattering reference 
plane, the longitude of the ascending node $\Omega$ can be estimated to be $16^o\pm6$.
 Coincidentally, the observation of HD189733b over half an orbital period indicates
\citep{knut} the peak hemisphere-integrated brightness to occur $16^o\pm6$ before opposition.
 Similar type of observation along with the polarimetry data for other planets may decide if
 there is any correlation. Taking the inclination angle $i=85^o.76$  as determined by transit
 method \citep{winn07}, I obtain the best fit by setting $\Omega=20^o$ and the modal grain
 diameter $d_0=0.8 \mu m$. The theoretical model alongwith the observed data is presented in
 figure~1. Since the inclination angle of the planet is nearly $90^o$,
 polarization is zero at the transit $(\Lambda=180^o)$ and at the secondary eclipse
 $(\Lambda=0^o)$. These are the positions when the planet's night and day sides are turned
 towards the observer. The polarization peaks near $\Lambda=90^o$ because polarization of
 light that is singly scattered is the largest for a scattering angle $\pi-\Lambda=90^o$  
and in a sufficiently thin medium, single scattering is favoured over multiple scattering. 
 In a non-circular orbit, the peak polarization shifts towards the longitude of pericentre
$\omega$ \citep{sen06}. The observed data for Q allows an eccentricity as high as e=0.06 
 so that $e\cos\omega=0.001$ for $\omega=89^o$ which is consistent with the time delay of the 
secondary eclipse \citep{knut}. However, more
 stringent limit is needed for $e\sin\omega$ in order to constrain the eccentricity
\citep{winn07}.  Contrary to the claim \citep{berd} that the orbital inclination angle
 greater than $90^o$ can be detected  through polarization, I do not find any change in 
the polarization profile if  the inclination is taken to be $90^o+i$.  At small values
 of x and the oblateness, the higher harmonic coefficients of the Stokes parameters
 are expected to be negligible as compared to the second one for Mie scattering.
 The ratio  of the second harmonic coefficients is equal to
 $(1+\cos 2i)/2\cos i$ and hence Fourier transformation of more accurately measured polarization
 data may provide the inclination angle of any extrasolar planet, transits as well as
 non-transits.

 Unlike the case of Brown Dwarfs which needs nonspherical photosphere to make incomplete 
cancellation of polarization when integrated over the disk, a perfectly spherical planet
can yield non-zero polarization because the illumination of the planetary surface by the
star-light does not cover the entire disk except when the planetary phase angle is $0^o$ or
$180^o$.  The rotation induced oblateness or tidal distrotion of the planetary disk introduces
additional asymmetry which increases the amount of polarization.  
For a non-spherical planetary disk, the shape of the planet could also affect the time variation
of the polarization. However, as estimated in section~\ref{obl}, the rotation induced oblateness
of the tidally locked planet HD 189733b is too small to yield any significant effect on the
polarization profile as comapred to the case of a perfectly spherical geometry. In fact, the
estimated oblateness of the planet increases the polarization by about twice its value 
calculated by considering spherical geometry. Therefore, the observed high peak amplitude of
polarization cannot be achieved by considering only non-spherical photosphere. Consequently,
the presence of dust in the photosphere of the planet becomes essential in order to explain the
observed polarization.  Finally, it remains to be verified if the inclusion of forsterite 
dominated thin cloud can give rise to a flux ratio of about $10^{-4}$ in the blue band. 

\section {Conclusion}

  The relatively high peak amplitude of the detected polarization and the large $1-\sigma$
errors in the data makes the reported polarization tentative and it remains to be confirmed
by further polarimetric observations with more accuracy. Until then, any interpretation
of the observation is only speculative. However, if confirmed, the reported polarization
would indicate the presence of highly reflecting species in the uppermost atmosphere
of HD 189733b that makes the B-band albedo much higher than the present theoretical
estimation. Rayleigh scattering of $H_2$ and $He$ may although make the flux ratio of
the planet and the star to reach the required order of magnitude ($10^{-4}$), it is 
unlikely that Rayleigh scattering would give rise to sufficient amount of 
polarization unless an unusually large radius of the planet is assumed.  
 On the other hand, recent observation \citep{pont} of an almost featureless transmission
 spectrum between 550 and 1050 nm with no indication of the expected alkaline absorption
 features suggests the presence of a haze of sub-micron particles in the upper atmosphere 
of HD 189733b. The important message conveyed by the present work is that the detected high amount
of polarization from HD 189733b, if correct, strongly supports the presence of a thin cloud
layer with sub-micron size grains in the visible atmosphere of the planet. 

\acknowledgments
I am thankful to Adam Burrows for kindly providing the theoretical temperature-pressure
profiles of the planet HD 189733b, to  
S. V. Berdyugina and D. M. Fluri for hospitality at ETH, Zurich, Switzerland
and to two anonymous referees for useful comments and suggestions.

\clearpage
\begin{figure}
\includegraphics[angle=0.0,scale=.90]{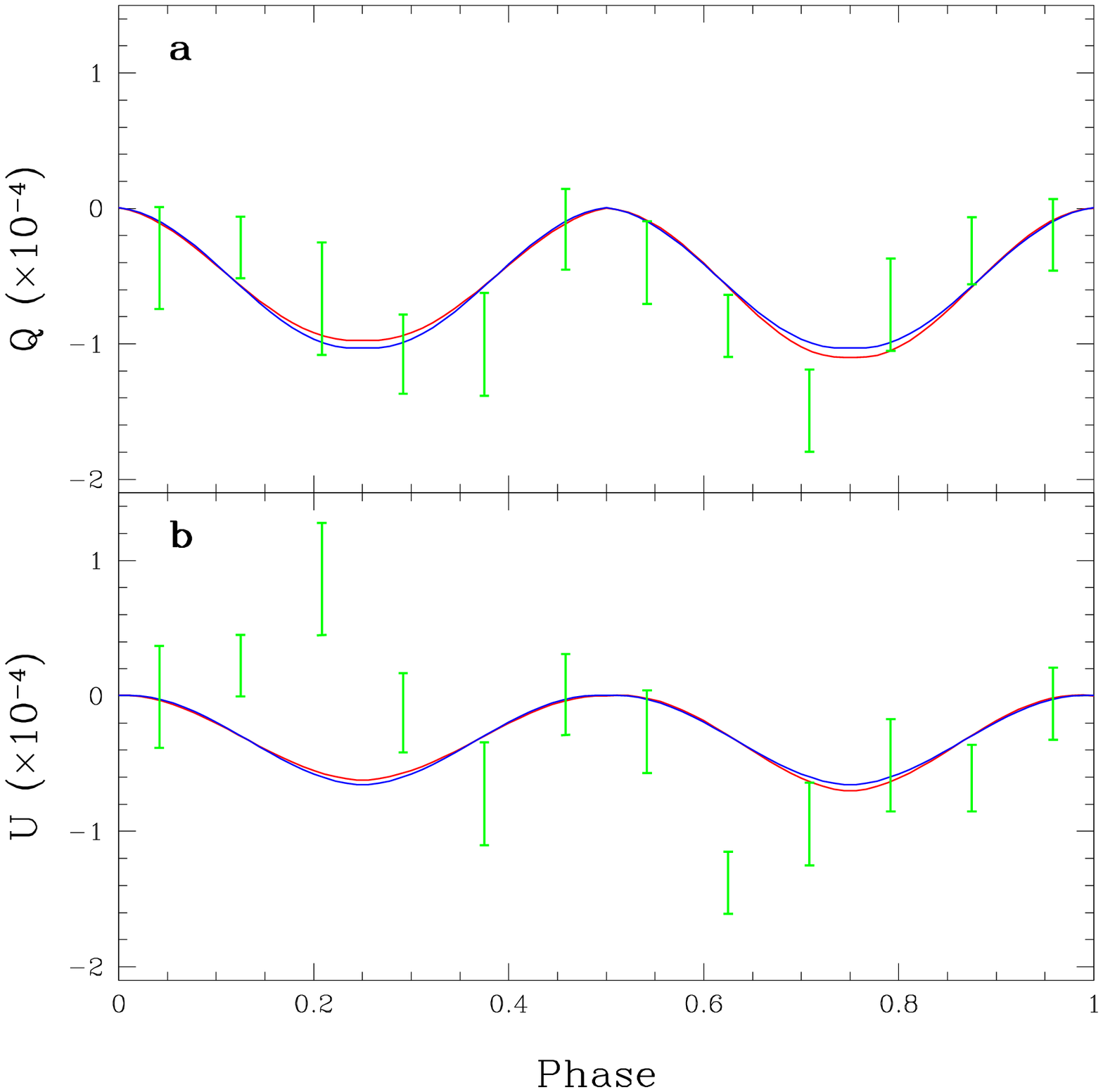}
\caption{
Best-fit models of the observed linear polarization at B-band of HD 189733b. The polarization
of HD 189733b expressed as the normalized and disk integrated Stokes 
Q (panel a) and U (panel b) for circular orbit (blue) and for elliptical orbit (red) with
eccentricity 0.06 and the longitude of pericentre at $89^o$.  The observed data,
re-binned for equal phase intervals, are presented by error-bars (green).  Q and U are on
the scale of $10^{-4}$.
\label{fig1}}
\end{figure}

\end{document}